\documentstyle[12pt]{article}
\title{\bf A Dual Gauge Model with Confinement }
\author{
G.A. Kozlov\\
\em Bogoliubov Laboratory of Theoretical Physics,\\
\em Joint Institute for Nuclear Research,\\
\em 141980 Dubna, Moscow Region, Russia\\
{M. Baldicchi}\\
\em Dipartimento di Fisica dell'Universit\`{a}-Milano,\\
\em Istituto Nazionale di Fisica Nucleare,\\
\em Sezione di Milano-Via Celoria 16,\\
\em 20133 Milano, Italy}

\begin{document}
\date{}
\maketitle
\begin{abstract}

{\small We reformulate the dual gauge model of the
long-distance Yang-Mills theory in terms of the two-point Wightman
functions formalism. In the flux-tube scheme of Abelian
dominance and monopole condensation, the analytic expressions of
both monopole- and dual gauge boson-fields propagators are
obtained. Formally, new features of higher order quadratic
equations for a monopole field are given. Finally, we show how
the rising-type potentials in the static system of color charges
are naturally derived.\\

 }
\end{abstract}

\section{ Introduction}
\setcounter{equation}{0}

There is a general statement that the color confinement is
supported by the idea that the vacuum of quantum Yang-Mills
(Y-M) theory is realized by a condensate of monopole-antimonople
pairs [1]. In such a vacuum the interacting field between two
colored sources located in $\vec{x}_{1}$ and $\vec{x}_{2}$ is
squeezed into a tube whose energy $E_{tube}\sim
\vert\vec{x}_{1}-\vec{x}_{2}\vert$. This is a complete dual
analogy to the magnetic monopole confinement in the type II
superconductor. Since there is no monopoles as classical
solutions with finite energy in a pure Y-M theory it has been
suggested by 't Hooft [2] to go into the Abelian projection
where the gauge group SU(2) is broken by a suitable gauge
condition to its (may be maximal) Abelian subgroup U(1).
It is proposed that the interplay between a quark-antiquark pair
is analagous to the interaction between a monopole-antimonopole
pair in a superconductor.

In fact, the topology of Y-M SU(N) manifold and that of its
Abelian subgroup $[U(1)]^{N-1}$ are different, and since any such gauge
is singular, one might introduce the string by performing the
singular gauge transformation with an Abelian gauge field
$A_{\mu}$ [3]
\begin{eqnarray}
\label{e1.1}
A_{\mu}(x)\rightarrow
A_{\mu}(x)+\frac{g}{4\pi}\,\partial_{\mu}\Omega(x)\ ,
\end{eqnarray}
where $\Omega(x)$ is the solid angle subtended by the closed
space-like curve described by the string at any point
$x=(x^0,x^1)$, and $g=2\pi/e$ is responsible for the magnetic flux
inside the string, $e$ being the Y-M coupling constant. Here, we
choose
a single string in the
two-dimensional (2d) world sheet $y_{x}(\tau,\sigma)$, for
simplicity. Obviously, the Abelian field-strength tensor
$F_{\mu\nu}^{A}=\partial_{\mu}A_{\nu}-\partial_{\nu}A_{\mu}$
transforms as
$$ F_{\mu\nu}^{A}(x)\rightarrow
F_{\mu\nu}^{A}(x)+\tilde {G}_{\mu\nu}(x)\ ,$$
where a new term
$$ \tilde
{G}_{\mu\nu}(x)=\frac{g}{4\pi}\,\lbrack\partial_{\mu},\partial_{\nu}\rbrack \,
\Omega(x)\ , $$
is valid on the world sheet only [4]
$$ \tilde {G}_{\mu\nu}(x)=\frac{g}{2}\,\epsilon_{\mu\nu\alpha\beta}\,\int\int d\sigma\,
d\tau\,
\frac{\partial (y^{\alpha},y^{\beta})}{\partial (\sigma
,\tau)}\,\,
\delta_{4} \lbrack x-y(\sigma ,\tau)\rbrack \ .$$
Formally, a gauge group element, which transforms a generic SU(N)
connection onto the gauge fixing surface in the space of
connections, is not regular everywhere in spacetime. The
projected (or transformed) connections contain topological
singularities (or defects). Such a singular transformation
(\ref{e1.1}) may form the worldline(s) of magnetic monopoles.
Hence, this singularity leads to the monopole current
$J_{\mu}^{mon}$. This is a natural way of the transformation
from the Y-M theory to a model dealing with Abelian fields.
A dual string is nothing but a formal idealization of a magnetic
flux tube in the equilibrium against the pressure of
surrounding superfluid (Higgs-like field) which it displaces
[5,6].

Recent lattice results [7] give the promised picture that the
monopole degrees of freedom can
indeed form a condensate responsible for the confinement. The
expression for the static heavy quark potential, using an
effective dual Ginzburg-Landau model [8], has been presented in
[9]. In the paper [10], an analytic approximation to the dual
field propagator without sources and in the presence of quark
sources, and an expression for the static quark-antiquark
potential were established.

The aim of this paper is to consider the model in 4d based on
the dual description of a long-distance Y-M (LDY-M) theory which
shows some kind of confinement. We study the model of Lagrangian
where the fundamental variables are an octet of dual potentials
coupled minimally to three octets of monopole (Higgs) fields.
The dual gauge model is studied at the lowest order of the
perturbative series using the canonical quantization. The basic
manifestation of the model is that it generates the equations of
motion where one of them for the scalar Higgs field looks like
as a dipole-like field equation. The monopole fields
obeying such an equation are classified by their two-point
Wightman functions (TPWF). In the classical level there is some
intersection with the Froissart model [11] containing the scalar
field satisfying the equation of the fourth order.
In the scheme presented in this work, the flux distribution in
the tubes formed between two heavy color charges is understood
via the following statement: the Abelian monopoles are excluded
from the string region while the Abelian electric flux is
squeezed into the string region.

In Sec. 2, we introduce the essence of the dual gauge Higgs
model and the classical solutions. In our model there are the
dual gauge field $\hat{C}_{\mu}^{a}(x)$ and the monopole field
$\hat{B}_{i}^{a}(x)$ ($i=1,..., N_{c}(N_{c}-1)/2$; $a$=1,...,8
 is a color index
) which
are relevant modes for infrared behaviour. The local coupling of
the $\hat{B}_{i}$-field to the $\hat{C}_{\mu}$-field provides
the mass of the dual field and, hence, a dual Meissner effect.
 Although
$\hat{C}_{\mu}(x)$ is invariant under the local transformation
of $U(1)^{N_{c}-1}\subset SU(N_{c})$,
$\hat{C}_{\mu}=\vec{C}_{\mu}\cdot\vec{H}$ is an $SU(N_{c})$-gauge
dependent object and does not appear in the real world alone
($N_{c}$ is the number of colors and $\vec{H}$ stands for the
Cartan superalgebra). The commutation relations, TPWF and
Green's functions as well-defined distributions in the space
$S(\Re^{d})$ of complex Schwartz test functions on $\Re^{d}$
will be defined in Sec. 3. In Sec. 4, we study the monopole- and dual
gauge-
field propagations. In Sec. 5, we obtain the asymptotic
transverse behaviour of both the dual gauge field and the
color-electric field. The analytic expression for the static
potential is obtained in Sec. 6. Sec. 7 contains the
discussion and conclusions.

\section{A dual Higgs gauge model and classical solutions}
\setcounter{equation}{0}

The dual description of the LDY-M theory is simply understood by
switching on the dual gauge field $\hat{C}_{\mu}(x)$ and the
three
scalar octets  $\hat{B}_{i}(x)$ (necessary to give mass to all $C_{\mu}^{a}$
and carrying color magnetic charge) in
the Lagrangian density (LD) $L$ [12]
\begin{eqnarray}
\label{e2.1}
L=2\,Tr\left [
-\frac{1}{4}\hat{F}^{\mu\nu}\hat{F}_{\mu\nu}+\frac{1}{2}\left
(D_{\mu}\hat{B}_{i}\right )^2\right ] - W\left
(\hat{B}_{i}\right )\ ,
\end{eqnarray}
where
$$ \hat{F}_{\mu\nu}=\partial_{\mu}\hat{C}_{\nu}-\partial_{\nu}\hat{C}_{\mu}-
ig\lbrack\hat{C}_{\mu},\hat{C}_{\nu}\rbrack\ ,$$
$$D_{\mu}\hat{B}_{i}=\partial_{\mu}\hat{B}_{i}-ig\,\lbrack\hat{C}_{\mu},\hat{B}_
{i}\rbrack\ ,$$
$\hat{C}_{\mu}$ and $\hat{B}_{i}$ are the SU(3) matrices, g is
the gauge coupling constant of the dual theory. The Higgs fields
develop their vacuum expectation values (v.e.v.)
$\hat{B}_{{0}_{i}}$ and the Higgs potential $W(\hat{B}_{i})$
has a minimum at $\hat{B}_{{0}_{i}}$. The v.e.v.
$\hat{B}_{{0}_{i}}$ produce a color monopole generating current
confining the electric color flux. It is known, the LD
(\ref{e2.1}) can generate classical equations of motion
carrying a unit of the $z_{3}$ flux confined in a narrow tube along
the $z$-axis (corresponding to quark sources at
$z=\pm\infty$). This is a dual analogy to the Abrikosov [13]
magnetic vortex solution.

 As the next step we introduce the color ansatz
\begin{eqnarray}
\label{e2.4}
\hat{C}_{\mu}=\sum_{a}C_{\mu}^{a}\,\frac{1}{2}\lambda_{a}\ ,
\end{eqnarray}
where the vector potential $C_{\mu}^{a}$ is dual to an ordinary
vector potential in the Y-M theory, (1/2)$\lambda^{a}$ are generators of SU(3).
Following paper [12] in the sense of representing the quark
sources by the Dirac string tensor
$\tilde{G}_{\mu\nu}(x)$ having the same color structure as in
(\ref{e2.4}), one can arrive at a more suitable form of the LD
(\ref{e2.1})
\begin{eqnarray}
\label{e2.5}
L(\tilde{G}_{\mu\nu})=-\frac{1}{3}G_{\mu\nu}^2+4{\vert
(\partial_{\mu}-igC_{\mu} )
\phi\vert}^2+2(\partial_{\mu}\phi_{3})^2-W(\phi,\phi_{3})\ ,
\end{eqnarray}
where
$$ G_{\mu\nu}=\partial_{\mu}C_{\nu}-\partial_{\nu}C_{\mu}+\tilde{G}_{\mu\nu}\
, $$
and $\phi(x)$ and $\phi_{3}(x)$ denote the complex scalar monopole fields.
We choose the color structure for the QCD-monopole field
$\hat{B}_{i}$ (belonging to the fundamental representation of
 $SU_{c}(3)$) like in [12], and the effective potential stands
$$ W(B,\bar{B},B_{3})=\frac{2}{3}\lambda\{ 11 [ 2 (
B^2 + \bar {B}^2 - B_{0}^2 )^2 + ( B_{3}^2 -
B_{0}^2 )^2 ] $$
$$ +7 [ 2 ( B^2+
\bar{B}^2 ) + B_{3}^2 - 3 B_{0}^2 ]^2\}\ , $$
where
$$ \phi\equiv\phi_{1}=\phi_{2}=B_{1,2}-i\bar{B}_{1,2}\
,\,\,\,\,\,\,
\phi_{3}=B_{3}\ , $$
while $\lambda$ provides the weak couplings between the scalar
fields.

The dual gauge field $C_{\mu}$ satisfies the relation
$$ \partial_{\mu}C_{\nu}-\partial_{\nu}C_{\mu}=^{\ast}(\partial_{\mu}A_{\nu}-
\partial_{\nu}A_{\mu}) $$
in the absence of charges, and the duality transformation is
realized by interchanging the gluon field $A_{\mu}(x)$ and
$C_{\mu}(x)$. In the maximally Abelian gauge [14]
$A_{\mu}(x)=A^{a}_{\mu}(x) (\tau^a/2)$ behaves as the Abelian
gauge field $C_{\mu}(x)=A^{3}_{\mu}(x) (\tau^3/2)$ approximately
because the off-diagonal fields are suppressed by the gauge
transformation.
 The LD (~\ref{e2.5}~) is invariant under the
local gauge transformation of the dual gauge field $C_{\mu}$
$$ C_{\mu}(x)\rightarrow
C_{\mu}(x)- \partial_{\mu}\Lambda (x)$$
and the phase transformation of the QCD-monopole field
$$ \phi_{1,2}(x)\rightarrow\exp
[-i\,g\Lambda (x)]\,\phi_{1,2}(x)\ , $$
where $\Lambda(x)$ is
the real field in $S(\Re^3)$ at any fixed $x^0$.
 The local gauge
symmetry is spontaneously broken because of
${\langle\phi\rangle}_{0}\not= 0$ in (\ref{e2.5}). The
generating current of (\ref{e2.5}) is nothing but the monopole
current confining the electric color flux
$$ J^{mon}_{\mu}=4\,i\,g[\phi^{\ast}
(\partial_{\mu}-i\,g\,C_{\mu})\,\phi\,-\,\phi (\partial_{\mu} +
i\,g\,C_{\mu})\,\phi^{\ast} ]\ , $$
which enters into the equation of motion in the form
$$ \partial^{\nu}\,G_{\mu\nu}(x)=\frac{3}{2}\,J^{mon}_{\mu}(x)\
.$$
The formal consequence of the $J^{mon}_{\mu}$ conservation,
$\partial^{\mu}J^{mon}_{\mu}=0$, means that monopole currents
form closed loops.

To find a solution of this model, one can consider the
monopole-field as a solution where ${\langle B(x)\rangle
}_{0}\not= 0$, ${\langle\bar{B}(x)\rangle
}_{0}\not= 0$, ${\langle B_{3}(x)\rangle
}_{0}\not= 0$. We choose
$$ B(x)=b(x)\, +\,B_{0}\ ,
\bar {B}(x)=\bar{b}(x)\ ,
B_{3}(x)=b_{3}(x)\, +\,B_{0} $$
with the boundary conditions at large distances $\rho$ from  the
center of the flux tube with $\vartheta$ as an angle in the
cylindrical coordinates [12]
\begin{eqnarray}
\label{e2.16}
\vec{C}\rightarrow -\frac{e}{g\,\rho}\vec{e}_{\vartheta}\
,\phi\rightarrow B_{0}\exp{(i\vartheta)}\ ,B_{3}\rightarrow
B_{0}\,\,\,\,
as\,\,\, \rho\rightarrow\infty\ .
\end{eqnarray}
In terms of the fields $b(x), \bar{b}(x), b_{3}(x)$ and
$C_{\mu}(x)$ the LD (\ref{e2.5}) is divided into two parts
\begin{eqnarray}
\label{e2.17}
L=L_{1}\,+\,L_{2}\ ,
\end{eqnarray}
where $L_{1}$ in the lowest order of $g$ and $\lambda$ and with
the minimal weak interaction looks like
\begin{eqnarray}
\label{e2.18}
L_{1}=-\frac{1}{3}G^{2}_{\mu\nu}\,+\,4\left
[(\partial_{\mu}b)^2+(\partial_{\mu}\bar{b})^2+\frac{1}{2}(\partial_{\mu}b_{3})^2\right
]\, \cr
+\,m^2\,C^2_{\mu}-\frac{4}{3}\mu^2
(50b^2+18b_{3}^2)+8m\partial_{\mu}\bar{b}\,C_{\mu}\ .
\end{eqnarray}
Here, $m\equiv gB_{0}$ and $\mu\equiv\sqrt{2\lambda}B_{0}$ are
masses of the dual gauge field and the monopole field,
respectively. The remaining part of (\ref{e2.17}) turns out to
be

$$ L_{2}=8g\left
[(\partial_{\mu}\bar{b})C_{\mu}b-(\partial_{\mu}b)C_{\mu}\bar{b}\right ]+
4g^2 {(\partial_{\mu}C_{\nu}-\partial_{\nu}C_{\mu})}^2\cdot
(\bar{b}^2 +b^2 +2B_{0}b) $$
$$-\frac{4}{3}\lambda [ 25(b^4 +\bar{b}^4) +9b_{3}^4
+100B_{0}b\cdot (\bar{b}^2 +b^2 +B_{0}^2) + 28B_{0}(\bar{b}^2 +b^2
+bb_{3})\cdot b_{3} $$
$$ +36B_{0}b_{3}(b_{3}^2 +B_{0}^2) + 2b^2(25\bar {b}^2
+7b_{3}^2)
 -2B_{0}^3(50b +18 b_{3}) +56B_{0}^2
bb_{3}+14\bar{b}^2b_{3}^2  ] . $$
Let us consider the canonical quantization of (\ref{e2.18}). The
equations of motion are
$$ (\Delta^2 +\mu_{1}^2)\,b(x)=0\ ; $$

\begin{eqnarray}
\label{e2.21}
\Delta^2\,\bar{b}(x) + m(\partial\cdot C)=0\ ;
\end{eqnarray}

$$ (\Delta^2 +\mu_{2}^2)\,b_{3}(x)=0\ ;$$
\begin{eqnarray}
\label{e2.23}
 (\Delta^2 +m^2_{1})\,C_{\mu}(x) -\partial_{\mu} (\partial\cdot C) +12\,m\,\partial
_{\mu}\bar{b}(x)-\partial^{\nu}\tilde{G}_{\mu\nu}(x)=0\ ,
\end{eqnarray}
where $\mu_{1}^2=(50/3)\,\mu^2, \mu_{2}^2=12\,\mu^2,
m_{1}^2=3\,m^2 $. By taking the divergence of (\ref{e2.23}) one
can get
\begin{eqnarray}
\label{e2.24}
\Delta^2\,\bar{b}(x) - \frac{1}{9\,m} \partial^\mu\partial^\nu\tilde{G}_{\mu\nu}
(x)=0\ ,
\end{eqnarray}
while the formal solution of equation (\ref{e2.23}) looks like
$$ C_{\mu}(x)=\alpha\,\partial^\nu\tilde{G}_{\mu\nu}(x) -
\beta\,\partial_{\mu}\bar{b}(x)\ , $$
where $\alpha\equiv (3\,m^2)^{-1}$, $ \beta\equiv 4/m$.
We see that the dual gauge field is defined via the divergence
of the scalar field $\bar{b}(x)$ shifted by the divergence of
the Dirac string tensor $\tilde{G}_{\mu\nu}(x)$. At the same
time, we propose in the standard manner the Dirac string which
can be understood as a straight line connecting two objects with
opposite charges. For large enough $\vec{x}$, approaching such a
string, the monopole field is going to its v.e.v. while
 $C_{\mu}(\vec{x}\rightarrow\infty)\rightarrow 0$.
Hence,

$$ J_{\mu}^{mon}(\vec{x}\rightarrow\infty)\rightarrow
8\,m^2\,C_{\mu}\ . $$
For a very weak $C_{\mu}$-field one can say that in the $d=2h$
dimensions
\begin{eqnarray}
\label{e2.27}
\Delta^{2h}\,\bar{b}(x)\simeq 0\ ,\,\,\,\,\,\,\, h=2,3,...\ ,
\end{eqnarray}
but
$$ \Delta^2\,\bar{b}(x)\not= 0\ . $$
Here, the solutions of equation (\ref{e2.27}) obey locality,
Poincare covariance and spectral conditions, and look like the
dipole "ghosts" at h=2. In this model, the role of the dipole
field at d=4 is held in the d=2 dimensions by the simple pole field,
and the analogy of the behaviour between the d=2 and d=4 dimensions
can be found at the level of Wightman functions in the free
case, at least [15]. This is related to the model proposed in
[16] in a study of the Higgs phenomenon, from which the present
model is distinguished by the coupling to the dual gauge field
$C_{\mu}(x)$, and the gauge field strengh tensor is shifted by
the Dirac string tensor $\tilde{G}_{\mu\nu}(x)$. Thus, we obtain
that the massless scalar field $\bar{b}(x)$ occurs in the model
since the symmetry realizes such a way that the LD (\ref{e2.5})
is invariant under the local gauge transformations above
mentioned but ${\langle \bar{b}(x)\rangle}_{0}\neq 0$.

Our aim is to find Green's function of a scalar field
$\bar{b}(x)$ obeying Eq. (\ref{e2.27}). The propagator $\tau_{h}(x)$ is
defined via TPWF $W_{h}(x)$ in the d=2\,h-dimensions
\begin{eqnarray}
\label{e2.30}
\tau_{h}(x)={\langle
T\,\bar{b}(x)\,\bar{b}(0)\,\rangle}_{0}=\theta (x^0)\,W_{h}(x) +
\theta (-x^0)\,W_{h}(-x)\ ,
\end{eqnarray}
where
\begin{eqnarray}
\label{e2.31}
W_{h}(x)={\langle
\,\bar{b}(x)\,\bar{b}(0)\,\rangle}_{0}
\end{eqnarray}
is the distribution in the Schwartz space $S^{\prime}(\Re^{2h})$
of temperate distributions on $\Re^{2h}$ and obeys the
equation
\begin{eqnarray}
\label{e2.32}
\Delta^{2\,h}\,W_{h}(x)=0 \ .
\end{eqnarray}
The general solution of eq. (\ref{e2.32}) should be Lorentz
invariant and is given in the form [17,18,16] at h=2
\begin{eqnarray}
\label{e2.33}
W_{2}(x)=a_{1}\,\ln\frac{l^2}{-x_{\mu}^2+i\,\epsilon\,x^{0}} +
\frac{a_{2}}{x_{\mu}^2-i\,\epsilon\,x^{0}} + a_{3}\ ,
\end{eqnarray}
where $a_{i}$ (i=1,2,3) are the coefficients to be defined
later, while $l$ is an arbitrary parameter with dimension minus
one in mass units. Its origin becomes more transparent from [15,
19]. In fact, the first and the second terms in (\ref{e2.33})
are related to the scalar dipole field and the scalar pole
field, respectively [15]. In the first case, the solution is
\begin{eqnarray}
\label{e2.34}
W_{2}(x)=-i\,E_{2}^{-}(x)=\frac{1}{(4\,\pi)^2}\,\ln\frac{l^2}{-x_{\mu}^2
+i\,\epsilon\,x^{0}}\ ,
\end{eqnarray}
while for the second term the solution looks like
\begin{eqnarray}
\label{e2.35}
W_{2}(x)=-i\,D_{2}^{-}(x)=\frac{1}{(2\,\pi)^2}\,\frac{1}{-x_{\mu}^2
+i\,\epsilon\,x^{0}}\ .
\end{eqnarray}

\section{TPWF as classical distributions}
\setcounter{equation}{0}

Before going into the quantization procedure, let us briefly consider
 the classical distributions (in the sense of generalized
functions [20]) of TPWF and T-ordered TPWF for the
$\bar{b}(x)$-field at large $x_{\mu}^2$. The TPWF (\ref{e2.31})
at h=2 is provided by the distribution
$\theta(p^0)\,\delta^{\prime}(p^2)$ as [21]
\begin{eqnarray}
\label{e3.1}
W_{2}(x)\sim\int\,d_{4}p\,\,\theta(p^{0})\,\delta^{\prime}(p^2)\,\exp(-i\,p\,x)\sim
-\ln[-\tilde{\mu}^2\,x^2 + i\,\theta(x^{0})] \cr
=-\left [\,\ln\vert\tilde{\mu}^2\,x^2\vert +
i\,\pi\,sgn(x^{0})\,\theta(x^2)\right ]\ ,
\end{eqnarray}
where $\tilde{\mu}$ is an arbitrary parameter (the infrared
regularization parameter).
The distribution $\theta(p^0)\,\delta^{\prime}(p^2)$ in
(\ref{e3.1}) is uniquely defined only on the test functions
$f(p)\in S_{0}(\Re_{4})$ $(S_{0}(\Re_{4})=\{f(p)\in S(\Re_{4}),
f(p=0)=0\})$

$$ \int\,d_{4}p\,\,\theta(p^{0})\,\delta^{\prime}(p^2)\,f(p)=
\lim_{\nu^{2}\rightarrow 0}
\,\frac{\partial}{\partial\nu^2}\,\int\,d_{4}p\,\,\theta(p^{0})\,\delta^{\prime}
(p^2-\nu^2)\,f(p)\ , $$
$$ =\int_{\Gamma_{0}^{+}}\,\frac{d_{3}\vec{p}}{2p^{0}}\,\frac{1}{2\,(n\,p)}\,\left
[\frac{1}{(n\,p)} - (n\,\partial)\right ]\,f(p) $$
and for an arbitrary fixed time-like unit vector $n_{\mu}$ in the
case we choose $n_{\mu}=(1,\vec{0})$ from $V^{+}$ where
$$ V^{+}=\left\{x\in\Re\,:x^{0}> \vert x\vert\leq\left
[\Sigma_{j=1}^{3}(x_{j})^2\right ]^{1/2}\right\} $$
is an open upper light cone in the M-space. Under the dilatation
transformation $x\rightarrow \alpha x$ ($\alpha > 0$) the TPWF
(\ref{e3.1}) acquires the additional term
\begin{eqnarray}
\label{e3.4}
W_{2}(x)\rightarrow W_{2}(\alpha
x)=W_{2}(x)-\frac{1}{2\,(2\,\pi)^2}\,\ln\alpha .
\end{eqnarray}
It could be interpreted as a spontaneous symmetry breaking of
the dilatation invariance of (\ref{e2.27}). This is an important
point in the special role of the field $\bar{b}(x)$.

In general, the $\tau_{h}(x)$-function (\ref{e2.30}) is
a well-defined distribution in $S^{\prime}(\Re^{2h})$ and obeys
the $h$-ordered quadratic differential equation in the d-dimension
of space-time
$$ \Delta^{2h}\,{\tau_{h}}^{d}(x)=\delta_{d}(x)\ ,\,\,\,\,
h=2,3,... \ , $$

$$ \Delta^{2}\equiv \frac{\partial^2}{\partial\,{x^{2}}_{1}} + ...+
\frac{\partial^2}{\partial\,{x^{2}}_{m}}-
\frac{\partial^2}{\partial\,{x^{2}}_{m+1}} - ...-
\frac{\partial^2}{\partial\,{x^{2}}_{m+n}}\ ; $$

in the following cases [20,22]\\
i) even $d$ and $h\geq d/2$
\begin{eqnarray}
\label{e3.6}
{\tau_{h}}^{d}(x)={(-1)}^{d/2-1}\,\frac{\exp[(\pi/2)\,i\,n]}{4^{h}\,(h-d/2)!\,(h-1)!}\,
{\left [\,\tilde{\mu}^2\,P(x) + i\,\epsilon\right
]}^{-d/2+h} \cr
\cdot\ln\left [\,\tilde{\mu}^2\,P(x) + i\,\epsilon\right
]\ ,
\end{eqnarray}
where $P(x)=x_{1}^2 +...+ x_{m}^2 - x_{m+1}^2 - ...-
x_{m+n}^2;$\\
ii) even $d$ and $h<d/2$
\begin{eqnarray}
\label{e3.7}
{\tau_{h}}^{d}(x)={(-1)}^{h}\,\frac{\exp[(\pi/2)\,i\,n]}{4^{h}\,\pi^{d/2}\,(h-1)!}\,
{\left [\,\tilde{\mu}^2\,P(x) + i\,\epsilon\right
]}^{-d/2+h}\,\Gamma\left (\frac{d}{2}-h\right )\ .
\end{eqnarray}
In the case of $d=2h$-dimensions, expression (\ref{e3.6}) can be
more simplified as follows:
$$ {\tau_{h}}^{d=2h}(x)={(-1)}^{h-1}\,\frac{\exp[(\pi/2)\,i\,n]}{4^{h}\,(h-1)!}\,
\ln\,\left [\,\tilde{\mu}^2\,P(x) + i\,\epsilon\right ] \ . $$
For the following nearly realistic consideration of the
confinement occurrence, we are interested in the case i)
(\ref{e3.6}) which allows one to study highly singular objects
for the confinement-like picture. In view of that, for $h=2$ in
the M-space following Zwanziger [23] the Fourier inversion
$F[{\tau_{2}}^{M}(x)]$ of the distribution ${\tau_{2}}^{M}(x)$
looks like
\begin{eqnarray}
\label{e3.9}
 F\left [\,{\tau_{2}}^{M}(x)\right
]=\frac{\pi^2}{2}\,\frac{\partial}{\partial\,p^{\mu}}\,\left
[\frac {p^{\mu}\,\ln(l^2\,p^2 -
i\,\epsilon)}{(p^2+i\,\epsilon)^2}\right ]
\end{eqnarray}
and obeying the equation
$$ (p^2)^2\,F\left [{\tau_{2}}^{M}(x)\right
]= 1\ ,\,\,\,\,\, p^2= (p^{0})^2-\Sigma_{j=1}^{3}\,p_{j}^2\ , $$
where $l=\exp(\gamma_{E}-1/2)\cdot \,(2\,\tilde{\mu})^{-1}$
($\gamma_{E}$ stands for the Euler constant). Note that in
terms of weak derivatives the distribution (\ref{e3.9}) can be
rewritten as [15]
\begin{eqnarray}
\label{e3.11}
F\left [{\tau_{h}}^{d}(x)\right ]
=\lim_{\kappa^2\rightarrow 0}\,\left
[\frac{(-1)^{d/2}}{(p^2-\kappa^2+i\,\epsilon)^{d/2}}+i\,\pi^{d/2}\,\delta_{d}(p)\,
\ln\left (\frac{\kappa^2}{4\,\tilde{\mu}^2}\right )\right ]\ .
\end{eqnarray}
for any $h$ allowed in the dimension $d$.
The remaining case ii) (\ref{e3.7}) leads to the following
singular object:
$$ F\left [\,{\tau_{h}}^{d}(x)\right ]
=\lim_{\kappa\rightarrow 0}\,
\tilde{\mu}^{-d}\,\left
(-X^2+i\,\kappa\right )^{-h} \ , $$
where
$\tilde{\mu}^2\,X^2=p_{1}^2+...+p_{m}^2-p_{m+1}^2-...-p_{m+n}^2$.
In the space M one has
$$ F\left [\,{\tau_{h}}^{M}(x)\right ]
=\lim_{\kappa\rightarrow 0}\,
\tilde{\mu}^{-4}\,\left
(\frac{\tilde{\mu}^2}{-p^2+i\,\epsilon}\right )^{2}\ . $$
\section{A dual field propagator}
\setcounter{equation}{0}

For calculation of the coefficients $a_{1}$ and $a_{2}$ in
(\ref{e2.33}) one has to introduce invariant functions
$E_{2}(x)=E_{2}^{-}(x)-E_{2}^{-}(-x)$ and
$D_{2}(x)=D_{2}^{-}(x)-D_{2}^{-}(-x)$ [23] (see also formulae
(\ref{e2.34}) and (\ref{e2.35})) which define the commutator
written in the general form [16]
\begin{eqnarray}
\label{e4.1}
\left [\,\bar{b}(x),\bar{b}(0)\,\right ]=(2\,\pi)^2\,i\,\left [
4\,a_{1}\,E_{2}(x)+a_{2}\,D_{2}(x)\right ]\ ,
\end{eqnarray}
$$ E_{2}(x)=(8\,\pi)^{-1}\,sgn(x^0)\,\theta(x^2)\ ,$$
$$ D_{2}(x)=(2\,\pi)^{-1}\,sgn(x^0)\,\delta(x^2)\ . $$
In the space $S^{\prime}(\Re^{4})$ on $\Re^{4}$ the propagator
for the $\bar{b}(x)$-field looks like [16]
\begin{eqnarray}
\label{e4.4}
\tau_{2}(x)=a_{1}\left [
\,\ln\vert\,\tilde{\mu}^2\,x_{\mu}^2\,\vert+i\,\pi\,\theta(x_{\mu}^2)\right
]+a_{2}\,\left [\,x_{\mu}^{-2}+i\,\pi\,\delta(x_{\mu}^2)\right
]+a_{3}\ .
\end{eqnarray}
The coefficients $a_{1}$ and $a_{2}$ in (\ref{e4.4}) can be
fixed using the canonical commutation relations (CCR)
$$ \left [C_{\mu}(x),\pi_{C_{\nu}}(0)\right
]_{\vert_{x^{0}=0}}=i\,g_{\mu\nu}\,\delta^{3}(\vec{x})\ , $$
$$ \left [\,\bar{b}(x),\pi_{\bar{b}}(0)\right
]_{\vert_{x^{0}=0}}=i\,\delta^{3}(\vec{x})\ , $$
respectively. Here, the conjugate momenta $\pi_{C_{\mu}}(x)$ and
$\pi_{\bar{b}}(x)$ look like
$$ \pi_{\bar{b}}(x)=8\,\left
[\partial^{0}\,\bar{b}(x)+m\,C^{0}(x)\right ]\ , $$

\begin{eqnarray}
\label{e4.8}
\pi_{C_{\mu}}(x)=-\frac{4}{3}\,G_{0\mu}(x)\ .
\end{eqnarray}
The direct calculation leads to (see also Appendix):
$$ a_{1}=\frac{m^2}{12\,(2\,\pi)^2}\ , $$
$$ a_{2}=-\frac{1}{6\,(2\,\pi)^2}\ . $$
Restricting the $W_{2}(x)$ function (\ref{e2.33}) to only the
first term, one can obtain that $\tau_{2}(x)$ satisfies the
equation
\begin{eqnarray}
\label{e4.11}
(\Delta^{2})^{2}\,\tilde{\tau}_{2}(x)=i\,\delta_{4}(x)\ ,
\end{eqnarray}
where $\tilde{\tau}_{2}(x)=3(2/m)^{2}\,\tau_{2}(x)$.
The formal Fourier transformation in $S^{\prime}(\Re_{4})$ gives
\begin{eqnarray}
\label{e4.12}
\hat{\tau}_{2}(p)=weak\lim_{\tilde{\kappa}^{2}<< 1}\,
\frac{i}{3\,(2\,\pi)^4}\,\left\{m^2\left
[\frac{1}{(p^2-\kappa^2+i\,\epsilon)^2}+i\,\pi^2\,\ln\frac{\kappa^2}{\tilde{\mu}^2}\,
\delta_{4}(p)\right ] \right. \cr
\left.
-\frac{1}{2}\,\frac{1}{p^2-\kappa^{2}+i\,\epsilon}\right\}.
\end{eqnarray}
Here, $\kappa$ is a parameter of representation and not the
analogue of the infrared mass,
$\tilde{\kappa}^{2}\equiv\kappa^2/p^2$. To derive (\ref{e4.12}),
we used the well-known mathematical trick with the generalized
functions [20,16,15]
$$ weak\lim_{\tilde{\kappa}^{2}<<
1}\,\int\,d_{2h}p\,\exp(-i\,p\,x)\,\frac{(-1)^{h}}{(p^2-\kappa^2+i\,\epsilon)^{h}}=
$$
$$ =\frac{2\,i}{\Gamma(h)\,(4\,\pi)^{h}}\{\ln2-\gamma_{E}-\ln(\kappa\sqrt{-x_{\mu}^2
+i\,\epsilon})$$
$$ + O
[-\kappa^2\,x_{\mu}^2,-\kappa^2\,x_{\mu}^2\,\ln(\kappa\,\sqrt{-x_{\mu}^2+i\,\epsilon})
]\}\ . $$
To go into the structure of the dual $C_{\mu}$-field propagator,
we need to define the general form of the commutation relation
$[\,C_{\mu}(x),C_{\nu}(y)\,]$. To do this procedure let us
consider the canonical conjugate pair
$\{C_{\mu},\pi_{C_{\nu}}\}$ (for the LD (\ref{e2.18})) where
$\pi_{C_{\mu}}(x)$ results from (\ref{e4.8}). The consequent
CCR looks like
\begin{eqnarray}
\label{e4.14}
\left
[\frac{4}{3}C_{\mu}(x),\partial_{\nu}C_{0}(0)-\partial_{0}C_{\nu}(0)-g_{0\nu}
(\partial\cdot C(0))+\Delta_{0\nu}(0)
\right
]_{\vert_{x^{0}=0}}=ig_{\mu\nu}\delta^{3}(\vec{x}),
\end{eqnarray}
where $\Delta_{\mu\nu}(x)=g_{\mu\nu}\,(\partial\cdot
C(x))-\tilde{G}_{\mu\nu}(x)$ tends to zero as $x\rightarrow
0$ and the Dirac string tensor $\tilde{G}_{\mu\nu}(x)$ obeys the
equation
$$ \left (\Delta^2 + M^2\right )\,\tilde{G}_{\mu\nu}(x)=0\
,\,\,\,\, as\,\, x\rightarrow 0 . $$
Here, the equations of motion (\ref{e2.21}) and (\ref{e2.24})
were used, and $M=3m$. Obviously, the following form of the
free $C_{\mu}$-field commutator (see Appendix)
\begin{eqnarray}
\label{e4.16}
\left
[\,C_{\mu}(x),C_{\nu}(0)
\right
]=ig_{\mu\nu}\left [\,\xi\,m_{1}^2\,E_{2}(x)+c\,D_{2}(x)\right ]\ ,
\end{eqnarray}
ensures the CCR (\ref{e4.14}) at large $x_{\mu}^2$. Here, both
$\xi$ and $c$ are real arbitrary numbers. The TPWF for
$C_{\mu}(x)$ stands as (see also [15,19])
\begin{eqnarray}
\label{e4.17}
w_{\mu\nu}(x)={\langle C_{\mu}(x)C_{\nu}(0)\rangle}_{0}=ig_{\mu\nu}\left
[\xi m_{1}^2 E_{2}^{-}(x)+cD_{2}^{-}(x)+c_{1}F_{2}^{-}(x)+c_{2}\right
],
\end{eqnarray}
where
$F_{2}^{-}(x)={\langle\,\alpha (x)\,\alpha (0)\,\rangle}_{0}=i\,x^2$
is TPWF for the harmonic field $\alpha(x)$, $c_{1}$ and
$c_{2}$ are arbitrary real numbers. In fact, the third and the
fourth terms in (\ref{e4.17}) did not appear in (\ref{e4.16})
due to the properties of the $F_{2}^{-}(x)$-function, namely,
$F_{2}(x)= F_{2}^{-}(x)-F_{2}^{-}(-x)=0$ and the triviality,
respectively.
One can easily verify that (\ref{e4.14}) and (\ref{e4.16})
can be used to derive the
following requirement for $\xi$
\begin{eqnarray}
\label{e4.18}
\xi=\frac{3}{4}-4\,c\ .
\end{eqnarray}
The free dual gauge field propagator in $S^{\prime}(\Re_{4})$ in
any local covariant gauge can be assumed as
\begin{eqnarray}
\label{e4.19}
\hat{\tau}_{\mu\nu}(p)=\int\,d^{4}x\,\exp(i\,p\,x)\,\tau_{\mu\nu}(x)
\cr
=i\,\left [g_{\mu\nu}-\left (1-\frac{1}{\zeta}\right
)\,\frac{p_{\mu}\,p_{\nu}}{p^2+i\,\epsilon}\right ]\cdot\left
[\xi\,m_{1}^2\,\hat{t}_{1}(p)+c\,\hat{t}_{2}(p)\right ]\ ,
\end{eqnarray}
where
$$ \tau_{\mu\nu}(x)=\frac{i\,g_{\mu\nu}}{(4\,\pi)^2}\,\left
[\,\xi\,m_{1}^2\,\ln(-\tilde{\mu}^2\,x_{\mu}^2+i\,\epsilon
)+\frac{c}{x_{\mu}^2+i\,\epsilon}\right ]\ ; $$

$$ \hat{t}_{1}(p)=weak\,\lim_{\tilde{\kappa}^2 <<1}\,\left
[\frac{1}{(p^2-\kappa^2+i\,\epsilon)^2} +
i\,\pi^2\,\ln\left(\frac{\kappa{^2}}{\tilde{\mu}^2}\right
)\,\delta_{4}(p)\right ]\ ; $$

$$ \hat{t}_{2}(p)=weak\,\lim_{\tilde{\kappa}^2 <<1}\, \frac{1}{2}\,
\frac{1}{(p^2-\kappa{^2}+i\,\epsilon)} \ . $$
The gauge parameter $\zeta$ in (\ref{e4.19}) is a real number.
According to Eq. (\ref{e4.11}), the following requirement on
Green's function $\tau_{\mu\nu}(x)$
$$ (\Delta^{2})^{2}\,\tilde{\tau}_{\mu\nu}(x)=i\,\delta_{4}(x)\
,$$
leads to that a constant $c$ has to be equal to zero
where $\tilde{\tau}_{\mu\nu}(x)=\tau_{\mu\nu}(x)/(\xi\,m_{1}^2)$.

At the end of this section, let us consider the v.e.v.
\begin{eqnarray}
\label{e4.24}
V_{\mu}(x)={\left\langle\,\left [\,J_{\mu}^{mon}(x),\bar{b}(0)\right
]\right\rangle }_{0}\ ,
\end{eqnarray}
where the monopole current expressed in terms of the $\bar{b}$-field
and $\tilde{G}_{\mu\nu}$ Dirac tensor is
$$ J_{\mu}^{mon}(x)=\frac{2}{3}\,\left [
\frac{3}{m}\,\partial_{\mu}{\Delta}^2\,\bar{b}(x)-\alpha\,{\Delta}^2\,\partial^{\nu}
\tilde{G}_{\mu\nu}(x)-\partial^{\nu}
\tilde{G}_{\mu\nu}(x) \right ]\ . $$
Using the commutation relation (\ref{e4.1}) and the well-known
relation between the following distributions like
$\Delta^2\,sgn(x^0)\,\theta(x^2)=4\,sgn(x^0)\,\delta(x^2)$, we
arrive at the formal expression for $V_{\mu}(x)$
\begin{eqnarray}
\label{e4.26}
V_{\mu}(x)=\frac{i\,m}{12\,\pi}\,\partial_{\mu}\left
[sgn(x^0)\,\delta (x^2)\right
]-\frac{2}{3}{\left\langle\,(\alpha\,\Delta^2-1)\,\left
[\partial^{\nu}\tilde{G}_{\mu\nu}(x),\bar{b}(0)\right
]\right\rangle}_{0}\ .
\end{eqnarray}
It is easily seen that the requirement of the Goldstone
theorem for occurrence of a $\delta(p_{\mu}^2)$ term in the
Fourier transformation of $V_{\mu}$ is satisfied due to the
presence of the first term in (\ref{e4.26}) which gives the
Fourier transformed term $\sim p_{\mu}\,sgn(p^0)\,\delta(p^2)$.
The second term in (\ref{e4.26}) comes from the dual string
effect.

\section{ The LDY-M solution for a gauge field}
\setcounter{equation}{0}

In this section, we consider the approximate topological solution
for the dual gauge field in the LDY-M theory. The equations of
motion are
\begin{eqnarray}
\label{e5.1}
 \partial^{\nu}\,G_{\mu\nu}=6\,i\,g[\phi^{\ast}
(\partial_{\mu}-i\,g\,C_{\mu})\,\phi\,-\,\phi (\partial_{\mu} +
i\,g\,C_{\mu})\,\phi^{\ast} ]\ ;
\end{eqnarray}
\begin{eqnarray}
\label{e5.2}
(\partial_{\mu}-i\,g\,C_{\mu})^2\,\phi=\frac{2}{3}\,\lambda\,(32\,B_{0}^2-25\,{\vert
\,\phi\,\vert}^2-7\,\phi_{3}^2\,)\,\phi\ .
\end{eqnarray}
At large distances Eq. (\ref{e5.2}) transforms into the
following one:
$$ (\partial_{\mu}-i\,g\,C_{\mu})^2\,\phi=\frac{50}{3}\,\lambda\,(\,B_{0}^2-\,{\vert
\,\phi\,\vert}^2\,)\,\phi\ . $$
To get the solution, let us do the polar decomposition of the monopole field
$\phi(x)$ using new scalar variables
 $\chi(x)$ and $f(x)$
$$ \phi(x)=\frac{1}{\sqrt{2}}\,\exp(i\,f(x))\,[\,\chi(x)+B_{0}\,]\
. $$
Then, the equation of motion (\ref{e5.1}) transforms into the following
one:
\begin{eqnarray}
\label{e5.5}
\partial^{\nu}\,G_{\mu\nu}=6\,g\,(\chi+B_{0})^2\,(g\,C_{\mu}-\partial_{\mu}\,f)\
.
\end{eqnarray}
This means that the $\bar{b}(x)$-field is nothing but a
mathematical realization of the "massive" phase $B_{0}\cdot f(x)$
at large enough distancese:
\begin{eqnarray}
\label{e5.6}
\bar{b}(x)=\frac{B_{0}}{2}\,S(x)\,f(x)+ \cr
m\int\,[\,2\,S(x)-1]\,C_{\mu}(x)
\,d\,x^{\mu} +\frac{1}{12\,m}\int\,[\,\Delta^2\,C_{\mu}(x)-\partial^{\nu}\partial_{\mu}\,
C_{\nu}(x)]
\,d\,x^{\mu}
\ ,
\end{eqnarray}
where $S(x)\equiv (1+\chi(x)/B_{0})^2$.
Now we have to define the flux as
\begin{eqnarray}
\label{e5.7}
\Pi=\int\,G_{\mu\nu}(x)\,d\,\sigma^{\mu\nu}\ ,
\end{eqnarray}
where $\sigma^{\mu\nu}$ is the 2d surface element in the
M-space. A substitution of $C_{\mu}(x)$ from (\ref{e5.5}) into (\ref{e5.7}) gives
\begin{eqnarray}
\label{e5.8}
\Pi=\frac{\alpha}{2}\,\oint_{\Gamma}\,S^{-1}(x)\,\partial^{\nu}\tilde{G}_{\mu\nu}(x)\,
d\,x^{\mu}
+\int\,\tilde{G}_{\mu\nu}(x)\,d\,\sigma^{\mu\nu}+\frac{1}{g}\oint_{\Gamma}\,
\partial_{\mu}\,f(x)\,d\,x^{\mu}\ ,
\end{eqnarray}
where $\Gamma$ means the large closed loop where the current
$\partial_{\mu}C_{\nu}-\partial_{\nu}C_{\mu}$ is canceled.
Integrating out over the loop $\Gamma$ in the third term in (\ref{e5.8}) is
nothing but the requirement that the phase $f(x)$ is varied by
$2\,\pi\,n$ for any integer number $n$ associated with the
topological charge [24] inside the flux tube. One can present
the fields $C_{\mu}(x)$ and $\phi(x)$ in the cylindrical
symmetry case using the radial coordinate $r$ (see also [24])
$$ \vec{C}\rightarrow\frac{\tilde{C}(r)}{r}\,\vec{e}_{\theta}\
,\,\,\,\,\, \phi\rightarrow\phi(r)\ , $$
and $f=2\,\pi\,n=n\,\theta$ with $\theta$ being the azimuth
around the z-axis. Thus, the field equation looks like
\begin{eqnarray}
\label{e5.10}
\frac{d^2\,\tilde{C}(r)}{d\,r^2}-\frac{1}{r}\,\frac{d\,\tilde{C}(r)}{d\,r}-
3\,m^2\,[\,3+2\,S(r)]\,\tilde{C}(r)+6\,n\,m\,B_{0}\,S(r)=0\ .
\end{eqnarray}
The following boundary conditions:
$$ \tilde{C}(r)=\frac{4\,n}{7\,g}\ ,\,\,\,\,\chi(r)=(\sqrt{2}-1)\,B_{0}\ ,\,\,
as\,\,r\rightarrow\infty\ ; $$
$$ \tilde{C}(r)=0\ ,\,\,\,\,\chi(r)=0\ ,\,\,
as\,\,r\rightarrow 0 $$
are conjugate with those in (\ref{e2.16}). At large enough
$r>>\mu^{-1}$ $((\mu=\sqrt{2\,\lambda}\,B_{0})^{-1}$ defines the
transverse dimension(s) of a monopole field around the tube)
Eq. (\ref{e5.10}) transforms into the following one:
$$ \frac{d^2\,\tilde{C}(r)}{d\,r^2}-\frac{1}{r}\,\frac{d\,\tilde{C}(r)}{d\,r}+
3\,g\,(4\,n-7\,g\,\tilde{C})\,B_{0}^{2}=0 $$
with the asymptotic transverse behaviour of its solution
$$ \tilde{C}(r)\simeq\frac{4\,n}{7\,g}-\sqrt{\frac{\pi\,m\,r}{2\,k}}\,e^{-k\,m\,r}\,
\left (\,1+\frac{3}{8\,k\,m\,r}\right )\ ,\,\,\,\,
k\equiv\sqrt{21}\ . $$
The field $\tilde{C}(r)$ grows rapidly when the radial distance
from the center of the flux tube $r<r_{0}\simeq 3~ fm$ and
approaches $4\,n/7\,g$ as soon as $r\geq r_{0}$.
Now it will be very instructive to clarify the singular
properties of the field $\bar{b}(x)$ related to the phase $f(x)$ by
means of (\ref{e5.6}). Since the phase is provided by the asimuth
$\theta$ around the z-axis,
$\vec{\nabla}\,f=(n/r)\,\vec{e}_{\theta}$. In the strong limit
of the mass $m$ of the $C_{\mu}$ field (as well as the mass of
the monopole field) at large distances
$$ \vec\nabla\times\vec{\nabla}\,\bar{b}=2\,\pi\,n\,B_{0}\,\delta(x)\,\delta(y)\,
\vec{e}_{z}\ , $$
where $\vec{e}_{z}$ is the unit vector along the axis $z$ and
the $\delta$-functions stand for the center of the flux tube
[24]. Hence, the real singular character of the field
$\bar{b}(x)$ is confirmed by its singular behaviour at the
center of the flux tube for nonzero monopole condensate. At the
end of this section we give the transverse distribution in $r$
of the color electric field $E_{z}(r)$ (see also [24])
$$ \hat{E}_{c}=\vec\nabla\times\vec{C}=\frac{1}{r}\,\frac{d\tilde{C}(r)}{d\,r}\,
\vec{e}_{z}\equiv E_{z}(r)\cdot\vec{e}_{z} $$
which has the following profile in the flux tube at large $r$
$$ E_{z}(r)=\sqrt{\frac{\pi\,m}{2\,k\,r}}\,\left
(k\,m-\frac{1}{2\,r}\right )\,e^{-k\,m\,r}\ . $$

\section{Static potential}
\setcounter{equation}{0}

In this section, we intend to obtain the confinement potential
in an analytic form for the system of interacting
static test charges of quark and
antiquark. Our statement
is based on the dual character of the field $C_{\mu}(x)$ to
a gluon field where $C_{\mu}(x)$ is just the interacting field
provided by the monopole field $\bar{b}(x)$ and the divergence
of $\tilde{G}_{\mu\nu}(x)$. We have found that $\bar{b}(x)$ plays the role of
the dipole-type field at $h=2$ (see Eq. (\ref{e2.27})). The mass
of the dual field $C_{\mu}(x)$ is nonzero and equal to
$m=g\,B_{0}$. It is assumed that the mass $m$ results from the
Higgs-like mechanism when the dual field $C_{\mu}(x)$ interacts
with the $\phi(x)$ field, namely an octet of dual potentials
$C_{\mu}$ coupled weakly with three octets of scalar fields $\phi_{\alpha}(x)
(\alpha =1,2,3)$.

According to the distribution (\ref{e3.11}), the first term in
the expansion for the static potential
$$ P_{stat}(r)=\int\,d_{3}\vec{p}\,e^{i\,\vec{p}\,\vec{r}}\,F\{{\tau_{h}}^{d}(x)\}_{\vert_
{p^{0}=0}} $$
in $\Re^{3}$ is a rising function with $r=\vert\vec{x}\vert$
[25,22]
\begin{eqnarray}
\label{e6.2}
P_{stat}(r)\sim\frac{1}{2^{2\,h}\,\pi^{3/2}}\,\frac{1}{(h-1)!}\,
 \Gamma (3/2-h)\,r^{2\,h-3}\ .
\end{eqnarray}
It is obvious that at $h\geq 2$ the distribution (\ref{e6.2})
increases with $r$ linearly ($h=2$) or faster ($h>2$). Let us
represent the distribution $r^{\sigma}$ as the Taylor series
around some regularization point $\sigma_{0}$
\begin{eqnarray}
\label{e6.3}
r^{\sigma}={\tilde{\mu}}^{-\omega}\,r^{\sigma_{0}}\,\left
[1+\omega\,\ln(\tilde{\mu}\,r)+\frac{1}{2}\,\omega^{2}\,\ln^{2}(\tilde{\mu}\,r)+...\right
]\ ,
\end{eqnarray}
where $\omega =\sigma -\sigma_{0}$ is an infinitesimal positive
interval at $\sigma\neq -d, -d-2,... $ The potential
(\ref{e6.2}) is simplified to [22]
$$ P_{stat}(r)\sim\frac{1}{8\,\pi}\,r\,[1+\ln (\tilde{\mu}\,r)+...]\
. $$
The Fourier transform $F\{r^{\sigma}\}$ of (\ref{e6.3}) into
$S^{\prime}(\Re_{d})$ for the limit $\omega\rightarrow 0$ leads
to the following singular distribution in the whole region of
the existence of the analytic function
$r^{\sigma}$ at $\sigma\neq -d, -d-2,...$:
$$ F\{r^{\sigma}\}=\left (\frac{4\,\pi}{p^2}\right
)^{(\sigma+d)/2}\,\pi^{-\sigma/2}\,\frac{\Gamma [(\sigma+d)/2]}{\Gamma (-\sigma
/2)}\ . $$
In general, the static potential is defined as
\begin{eqnarray}
\label{e6.6}
P_{stat}(r)=\lim_{T\rightarrow\infty}\,\frac{1}{T}\,A(r)\ ,
\end{eqnarray}
where the action $A(r)$ is given by the colour source-current
part of LD
$$L(p)=-\vec{j}_{\alpha}^{\mu}(-p)\,\hat{\tau}_{\mu\nu}(p)\,\vec{j}_{\alpha}^{\nu}(p)\
. $$
It is known that for such a system of heavy particles the
sources are given by a c-number current
$$ \vec{j}_{\alpha}^{\mu}(x)=\vec{Q}_{\alpha}\,g^{\mu
0}\,[\delta_{3}(\vec{x}-\vec{x_{1}})-\delta_{3}(\vec{x}-\vec{x_{2}})]
$$
with $\vec{Q}_{\alpha}=e\,\vec{\rho}_{\alpha}$ being the Abelian
color-electric charge of a quark while $\vec{\rho}_{\alpha}$ is
the weight vector of the SU(3) algebra: $\rho_{1}=(1/2,\,\sqrt{3}/6)$,
$\rho_{2}=(-1/2,\,\sqrt{3}/6)$, $\rho_{3}=(0,\,-1/\sqrt{3})$
[24]. Here, $\vec{x}_{1}$ and $\vec{x}_{2}$ are the position vectors
of a quark and an antiquark, respectively; the label
$\alpha$=(1,2,3) corresponds to the color electric charge. The
calculation of the potential (\ref{e6.6}) is most easily
performed by taking into account the Fourier transformed quark
current
\begin{eqnarray}
\label{e6.9}
 \vec{j}_{{\mu}_{\alpha}}(p)=2\,\pi\,\vec{Q}_{\alpha}\,g_{\mu
0}\,\delta (p^{0})\,\left (e^{-i\,\vec{p}\,\vec{x}_{1}}-e^{-i\,\vec{p}\,\vec{x}_{2}}
\right )\ ,
\end{eqnarray}
and by making use of the representation in the sense of
generalized functions [23]
\begin{eqnarray}
\label{e6.10}
weak \lim_{{\tilde{\kappa}}^2 <<1}\,\left
[\frac{1}{(p^2-\kappa^2+i\,\epsilon)^2}+i\,\pi^2\,\ln\frac{\kappa^2}{\tilde{\mu}^2}\,
\delta_{4}(p)\right ]= \cr
=\frac{1}{4}\frac{\partial^2}{\partial\,p^2}\,\frac{1}{-p^2-i\,\epsilon}\,\ln
\frac{-p^2-i\,
\epsilon}{\tilde{\mu}^2}=\frac{1}{2}\,\frac{1}{(p^2+i\,\epsilon)^2}\,\left
(5-3\,\ln\frac{-p^2-i\,
\epsilon}{\tilde{\mu}^2}\right )\ .
\end{eqnarray}
Due to the presence of the $\delta(p^{0})$-function (see
(\ref{e6.9}) in evaluating of the action $A(r)$, the
remaining 3-dimensional integral over $d_{3} p$ will be easily
calculated out using the instructive formula [20,15]
\begin{eqnarray}
\label{e6.11}
\int
d_{3}\vec{p}e^{i\vec{p}\vec{x}}p^{m}\ln^{\beta}p=\frac{1}{r^{m+3}}
\sum_{i=1}^{\beta}\frac{\Gamma(\beta+1)(-1)^{i}}{\Gamma(\beta-i+1)\Gamma(i+1)}
\left (\frac{d}{d\,m}\right )^{\beta-i}H_{m}\ln^{i}r,
\end{eqnarray}
where
$$ H_{m}=2^{m+3}\,\pi^{3/2}\,\frac{\Gamma [(m+3)/2]}{\Gamma (-m/2)}\
,\,\,\,\, m\neq -3,-5,...\ , $$
$r\equiv \vert\vec{x}\vert$, $p\equiv \vert\vec{p}\vert$,
$\vec{x}\in\Re^{3}$, $\vec{p}\in\Re_{3}$.

As a consequence of the dual field propagator (\ref{e4.19}) the
static potential (\ref{e6.6}) at large distances looks like
\begin{eqnarray}
\label{e6.13}
P_{stat}(r)=\frac{\vec{Q}^2}{16\,\pi}\left
\{\xi\,m^2\,r[5+6(A+\ln\tilde{\mu}\,r
)]+O\left(\frac{c}{r}\right )\right\} \ ,
\end{eqnarray}
where $A\equiv
1+\sqrt{\pi}-(1/2+2\,\sqrt{\pi}\,\gamma_{E})-(2\,\sqrt{\pi}+1)\,\ln2
< 0$ and the last term in (\ref{e6.13}) is just the positive
correction and not the analogue of the Coulomb part of the
potential due to the one-gluon exchange. Neglecting the last term in
(\ref{e6.13}) and taking into account formula (\ref{e4.18}),
one can conclude
\begin{eqnarray}
\label{e6.14}
P_{stat}(r)\simeq\frac{3\,\vec{Q}^2}{64\,\pi}
\,m^2\,r(-12.4+6\,\ln\tilde{\mu}\,r
) \ .
\end{eqnarray}
Hence, the string tension $a$ in $P_{stat}(r)=a\,r$ emerges as
\begin{eqnarray}
\label{e6.15}
a\simeq\frac{\alpha_{s}}{16}
\,m^2\,\left
(-12.4+3\,\ln\frac{\tilde{\mu}^2}{m^2}\right ),\,\,\,\,\, \tilde{\mu}
>9\,m\ ,
\end{eqnarray}
where $r$ in the logarithmic function in (\ref{e6.14}) has been
changed by the characteristic length $r_{c}\sim 1/m$ which
determines the transverse dimension of the dual field
concentration, while $\tilde{\mu}$ is associated with the
"coherent length" inverse and the dual field mass $m$ defines
the "penetration depth" in the type II superconductor.
For a typical value of the electroweak scale $\tilde{\mu}\simeq
250~GeV$ we get $a\simeq 0.20~GeV^2$ for the mass of the dual
$C_{\mu}$-field $m=0.6~GeV$ and $\alpha_{s}=e^2/(4\,\pi)$=0.37
obtained from fitting the heavy quark-antiquark pair spectrum
[26]. The value of the string tension (\ref{e6.15}) is quite
close to a phenomenological one (eg., coming from the Regge
slope of the hadrons).
Making the
formal comparison of the result obtained here in the analytic form
let us remind the analogue with the well-known expression of the
energy per unit length of the vortex in the type II superconductor
[27,9]
\begin{eqnarray}
\label{e6.16}
\epsilon_{1}=\frac{{\phi_{0}}^2\,m_{A}^2}{32\,\pi^{2}}\,\ln\left
(\frac{m_{\phi}}{m_{A}}\right )^2\ ,
\end{eqnarray}
where $\phi_{0}$ is the magnetic flux of the vortex, $m_{A}$ and
$m_{\phi}$ are penetration depth and the coherent length
inverse, respectively. On the other hand, the string tension in
Nambu's paper (see the first ref. in [1]) is given by
\begin{eqnarray}
\label{e6.17}
\epsilon_{2}=\frac{g_{m}^2\,m_{v}^2}{8\,\pi}\,\ln\left
(1+\frac{{m_{s}}^2}{{m_{v}}^2}\right )\ ,
\end{eqnarray}
with $m_{s}$ and $m_{v}$ being the masses of scalar and vector
fields and $g_{m}$ is a magnetic-type charge. It is clear from
 formula (\ref{e6.13}) that for a
sufficiently long string $r>>m^{-1}$ the $\sim r$-behaviour of
the static potential is dominant; for a short string $r<<m^{-1}$
the singular interaction provided by the second term in (\ref{e6.13})
becomes important if the average size of the monopole is even
smaller.
\section{Summary and discussion}
\setcounter{equation}{0}

We studied the dual gauge model of the long-distance Yang-Mills
theory in terms of two-point Wightman functions. The
quantization of the model has been provided by using
CCR, thus avoiding other methods (e.g., path integral use). We
intended to give our understanding of the confinement by making
use of nothing else but the well-known tools of quantum field
theory based on LD given in [12] as well as the
renormalization model and symmetry properties. Among the
physicists dealing with the models of interplay of a scalar
(monopole, Higgs) field with a dual vector (gauge) boson field,
where the vacuum state of the quantum Y-M theory is realized by a
condensate of the monopole-antimonopole pairs, there is a strong
belief that the flux-tube solution explains the scenarios of
color confinement. Based on the flux-tube scheme approach of
Abelian dominance and monopole condensation, we have obtained the
analytic expressions for both the monopole and  dual gauge
boson field propagators (\ref{e4.12}) and (\ref{e4.19}),
respectively, in $S^{\prime}(\Re_{4})$. These propagators lead to
a consistent perturbative expansion of Green's functions.
However, the Fourier transformation of the first term in TPWF (\ref{e2.33})
gives the occurrence of the $\delta^{\prime}(p_{\mu}^2)$-function.
This is a consequence of the nonunitarity of the translations,
and the spectral function with such a term gives an indefinite
metric [16]. We have found that the Goldstone theorem is valid in our
model in the form taking into account the Dirac's string
effect (\ref{e4.26}). In fact, we obtained that the
characteristic $\delta (p_{\mu}^2)$ term naturally appears in
the Fourier transformation of the v.e.v. of the commutator (\ref{e4.24})
 of the monopole current and the scalar local field $\bar{b}$.
In principle, a similar result has to be expected if one
replaces the $\bar{b}$-field in (\ref{e4.24}) by any product of
the gauge field $C_{\mu}$ and $\bar{b}$. We see that the fields
$b(x)$ and $b_{3}(x)$ receive their masses and the $\bar{b}(x)$
field in combination with $\partial^{\nu}\,\tilde{G}_{\mu\nu}(x)$ form
the vector field $C_{\mu}(x)$ obeying the equation of motion for
the
massive vector field with the
  mass $m=g\,B_{0}$. The solution of the $\bar{b}(x)$-field can
be identified as a "ghost"-like particle in the substitute
manner.

The occurrence of an arbitrary parameter $l$ in (\ref{e2.33}) and
(\ref{e2.34}) leads to breaking their covariance under the
dilatation transformation (\ref{e3.4}) and provides
spontaneous symmetry breaking of the dilatation invariance of
Eq. (\ref{e2.27}). The monopole condensation, formulated in the
framework of LDY-M model, causes the strong and long-range
interplay between heavy quark and antiquark, which gives the
confining force, through the dual Higgs mechanism.
We have obtained the analytic expression
for the static potential (\ref{e6.13}) at large distances.
 The form of this
potential grows linearly with the distance $r$ apart from
logarithmic correction. The latter comes from the second term in
the expression (\ref{e6.10}) (see also (\ref{e6.11}) ).

Making an analytic comparison of $\epsilon_{1}$ (\ref{e6.16})
and $\epsilon_{2}$ (\ref{e6.17}) with $a$ in (\ref{e6.15}),
one can conclude that we have obtained a similar behaviour of
the string tension $a$ to those in the magnetic flux picture of
the vortex and in the Nambu scheme, respectively, as well as
in the dual Ginzburg-Landau model [9].

Finally, it is to be noted that we have played the game with the
choice of the gauge group where the Abelian group appears as a
subgroup of the full Y-M gauge group. This is a very instructive
method of calculating the confinement potential in the static
limit in the analytic form. However, we understand that no real
physics can depend on such a choice. Now, there is a next step in
more formal consideration of the Y-M theory where it seems to be a
new mechanism of confinement [28,29].

\section{Acknowledgements}
G.A.K. is grateful to G.M. Prosperi for the kind hospitality at
the University of Milan where this work has partly been done.

\section{Appendix}
\setcounter{equation}{0}

The following relations [21,16,15] were used for calculation of
the coefficients $a_{1}$ and $a_{2}$ in (\ref{e4.1}):
$$ D_{2}(x^{0}=0,\vec{x})=0\ ,\,\,\,\,\,\,
\partial_{\mu}\,{D_{2}(x)}_{\vert_{x^{0}=0}}=g_{o\mu}\,\delta_{3}(\vec{x})\ ;$$
$$
(\Delta^{2})^{2}\,E_{2}(x)=\partial_{0}^{2}\,{E_{2}(x)}_{\vert_{x^{0}=0}}=
\partial_{0}\,{E_{2}(x)}_{\vert_{x^{0}=0}}=E_{2}(0,\vec{x})=0\
;$$
$$ \partial_{0}^{3}\,{E_{2}(x)}_{\vert_{x^{0}=0}}=8\,\pi\,g_{0\mu}\,\delta_{3}(\vec{x})\
,\,\,\,\,\,\, \Delta^2\,E_{2}(x)=D_{2}(x)\ .$$

\end{document}